\documentclass[pra,
twocolumn,
superscriptaddress,showpacs,floatfix,nofootinbib]{revtex4}
\usepackage{epsfig,amsmath,amsfonts,amssymb}
\usepackage{graphicx}
\usepackage[utf8]{inputenc}
\usepackage{boxedminipage}
\usepackage{epsfig,psfrag}
\usepackage{amsmath}
\usepackage{mathbbol, amsfonts}
\usepackage{lscape}
\usepackage{fancybox}
\usepackage{amsfonts,amssymb}
\usepackage{color}
\usepackage{subfig}
\newcommand{\beq}{\begin{equation}}
\newcommand{\eeq}{\end{equation}}

\newcommand{\beqa}{\begin{eqnarray}}
\newcommand{\eeqa}{\end{eqnarray}}
\def\ra{\rangle}
\def\la{\langle}

\usepackage{wrapfig}

\begin{document}
\title{Driving at the quantum speed limit: Optimal control of  a two-level system
}
\author{Gerhard C. Hegerfeldt}
\affiliation{Institut für Theoretische Physik, Universität Göttingen,
Friedrich-Hund-Platz 1, D-37077 Göttingen, Germany}

\begin{abstract}
A remarkably simple result is derived for the minimal time $T_{\rm min}$ required to drive a general initial state to a final target state by a Landau-Zener type Hamiltonian or, equivalently, by time-dependent laser driving. The associated protocol is also derived. A surprise arises for some states when the interaction strength is bounded by a constant $c$. Then, for large $c$, the optimal driving is of type bang-off-bang and for increasing $c$ one recovers the unconstrained result. However, for smaller $c$ the optimal driving can suddenly switch to bang-bang type. We discuss the notion of quantum speed limit time.
\end{abstract}
\maketitle

In various areas of physics the challenging task arises to change a given initial quantum state to a prescribed final target state in a controlled and optimal way, e.g. in quantum computation \cite{Mo4}, in quantum optics for fast population transfer \cite{Mup.4} or for Bose-Einstein condensates \cite{Mup3}.  One way to do this is to use an arbitrary slow change of the dynamical parameters and the adiabatic theorem
 \cite{Mu:BornFock1928}. However, often such an adiabatic approach may be too slow, and therefore various protocols have been devised to speed up the process. For example, the 'transitionless tracking algorithm' \cite{Berry1} adds a so-called counterdiabatic term to achieve adiabatic dynamics with respect to the original Hamiltonian in shorter time, while 'shortcuts to adiabadicity' (STA) \cite{Mup3,Chen} does not follow adiabatic states. For a comprehensive review of these and other approaches see Ref. \cite{Torr}.
An experimentally important requirement for such protocols is fidelity. Additional requirements may be small energy input, robustness, or that the target state is reached in the shortest time allowed by the available dynamics. Other approaches consider unitary time-development operators and aim to determine the optimal dynamics that leads from an initial $U(0)$ to a prescribed final propagator $U_F$ in minimal time \cite{Khaneja}. These approaches will not be further considered here. Ref. \cite{Carlini} used a variational approach to determine time optimal Hamiltonians.

A recent paper \cite{Mo} studied the experimental implementation of control protocols for the 'simplest non-simple quantum problem' \cite{Berry2} in which two states  are coupled by a Landau-Zener (LZ) type Hamiltonian of the form
\beq\label{2.1}
H= \Gamma(t)\sigma_3 + \omega(t) \sigma_1.
\eeq
where  $\Gamma$ corresponds to quasi-momentum in Ref. \cite{Mo}.
In optical driving of a two-level system a similar Hamiltonian applies, 
 with $\Gamma = -\Delta$, the detuning, and $\omega= \Omega$ the Rabi frequency.
In the experiments of Ref. \cite{Mo} an effective two-level system was realized by using Bose-Einstein condensates in an accelerated optical lattice.  
For time-independent $\omega$, numerical simulations \cite{Caneva,Mo} for special parameter values and for special initial and final groundstates indicate that the shortest possible time $T_{\rm min}$ is achieved by the 'composite pulse' protocol which represents a period with $\Gamma(t)\equiv 0$ preceded and followed by a $\delta$-like pulse of pulse area $\pi/4$. This was borne out by the experimental results of Ref. \cite{Mo}.

In this paper we  consider a two-level system with the LZ Hamiltonian of Eq. (\ref{2.1}) as well as a system driven by a time-dependent laser amplitude and detuning. We derive a remarkably simple expression, Eq.~(\ref{cos}), for the minimal time $T_{\rm min}$ required to change an arbitrary initial state to a final target state, under the condition that only interactions of the LZ type as in Eq. (\ref{2.1}) are considered. First the case is treated where the interaction strength $\Gamma(t)$ is not constrained in size and where $\omega(t)=$ const $= \omega$. In this case the  optimal protocol is the composite protocol, i.e. initial and final $\delta$-like pulses and in between a period with $\Gamma(t) \equiv 0$ of length $T_{\rm min}$, as in  the special case of the simulation in Refs. \cite{Mo,Caneva}. For initial and final states which are groundstates of the LZ Hamiltonian  the  pulse strengths are $\pm\pi/4$, while in general they are different. When an arbitrary $\omega(t)$ is allowed, the condition $|\omega(t)|\leq \omega_{\rm max}$ is imposed (otherwise $T_{\rm min}\to 0$ as $\omega \to \infty$). Then the optimal protocol is $\Gamma(t) \equiv 0$, $\omega(t) \equiv \omega_{\rm max}$ and again an initial and final $\delta$-like pulse.

Since arbitrarily large driving is a physical idealization we also consider the constrained case, $|\Gamma(t)| \leq c$. Surprisingly, now the optimal protocol depends on the value of $c$. For simplicity we consider for the constrained case only initial and final states  which are eigenstates of the LZ Hamiltonian, with $\Gamma=\gamma_{\rm in} = - \gamma$ and $\Gamma=\gamma_{\rm f} =  \gamma$, $ \gamma >0$. It is shown that for $ c > \omega^2/\gamma $ the optimal protocol is of bang-off-bang type, i.e. an initial and final period with $\Gamma(t)\equiv\pm c$ and in between a period with $\Gamma(t) \equiv 0$, 
 as expected from the unconstrained case, and for $c \to \infty$ one recovers the unconstrained result. However, if $c \leq \omega^2/\gamma $, the optimal protocol is of bang-bang type, and there is no time period with $\Gamma(t)\equiv 0$.

We also compare $T_{\rm min}$ with the quantum speed limit times $T_{\rm qsl}$ used in Refs. \cite{Mo} and \cite{Caneva}. It is shown that $T_{\rm min}$ is less or equal to $T_{\rm qsl}$ of Ref. \cite{Mo}, in fact it is equal to the latter when initial and final state are groundstates of the LZ Hamiltonian, but for more general states $T_{\rm min}$ can be less than $T_{\rm qsl}$ of Ref. \cite{Mo}.

{\em The control problem.} The aim is to find an optimal driving $\Gamma(t)$ such that the time-development operator $U_H(t,0)$ associated with $H$ in Eq. (\ref{2.1}) evolves an initial state $|\psi_{\rm in}\ra$ at time $t=0$ to (a multiple of) a final state $|\psi_f\ra$ at time $T$, i.e.
\beq \label{2.4a}
U_H(T,0) |\psi_{\rm in}\ra = \kappa\, |\psi_f\ra~,
\eeq 
and to find the minimal time $T_{\rm min}$. If  $|\psi_{\rm in}\ra$ and $|\psi_f\ra$ are normalized to 1,  $\kappa$ is a phase factor, otherwise it also contains the ratio of the normalization factors. To determine the minimal time required we use the Pontryagin maximum principle (PMP) \cite{PMP} which is explained further below.

As a consequence of  the Eulerian  rotation angles  any 
$U\in {\rm SU(2)}$  can be written in the form $U = \exp(-i\sigma_3\gamma/2) \exp(-i\sigma_1\beta/2) \exp(-i\sigma_3\alpha/2)$. Here it is convenient (but not necessary) to rewrite this. Writing $U_H(t,0) = U(t)\exp(i\sigma_1\pi/4)$, $\tau_3 \equiv  \gamma(t), \tau_1 \equiv  \beta(t) + \pi/2,$ and $\tau_2 \equiv  \alpha(t) $, one arrives at 
\beq \label{2.5}
U_H(t,0) = e^{-i \sigma_3 \tau_3(t)/2 }~ e^{-i \sigma_1 \tau_1(t)/2}~ e^{-i \sigma_2 \tau_2(t)/2}~,
\eeq
with as yet unkown functions $\tau_i(t)$.
This 
holds for any traceless  Hamiltonian. 
We now differentiate both sides of Eq.~(\ref{2.5}), equate the result with $\dot{U}_H =  -i\{\Gamma \sigma_3 +\omega \sigma_1\}U_H$, and  multiply by 
$e^{i \sigma _3\tau_3/2}$ from the left and by $ ~ e^{i \sigma_2 \tau_2/2}~e^{i \sigma_1 \tau_1/2} $from the right. This gives
\beqa
\dot\tau_3 \sigma_3 + \dot\tau_1 \sigma_1 + \dot\tau_2 e^{-i \sigma_1 \tau_1/2}\sigma_2 e^{i \sigma_1 \tau_1/2 }\nonumber\\
=2  e^{i \sigma _3\tau_3/2}  \{\Gamma\sigma_3 +\omega \sigma_1\} e^{-i \sigma_3 \tau_3/2 }.\nonumber
\eeqa
Using $e^{-i\sigma_1\tau_1/2}\sigma_2e^{i\sigma_1\tau_1/2} =\cos\tau_1 \sigma_2 + \sin \tau_1\sigma_3 $ etc. one obtains
\beqa
\dot\tau_3 \sigma_3 + \dot\tau_1 \sigma_1 + \dot\tau_2 (\cos\tau_1\,\sigma_2 + \sin\tau_1 \, \sigma_3)\nonumber\\
= 2\Gamma\, \sigma_3 +2 \omega (\cos\tau_3 \, \sigma_1 - \sin\tau_3 \, \sigma_2).
\eeqa
Since the $\sigma_i$'s are linearly independent this leads to a system of three equations for $\dot\tau_i$,
\beqa \label{2.9}
\dot \tau_1 &=& 2\omega\, \cos\tau_3\nonumber\\
\dot \tau_2 &=& - 2\omega\, \sin\tau_3/\cos\tau_1\\
\dot\tau_3  &=& 2\Gamma + 2\omega \sin\tau_3\, \sin\tau_1/\cos\tau_1~.\nonumber
\eeqa
 These equations can also be used for an engineering type approach by first prescribing a $\tau_3(t)$, then solving for $\tau_1(t)$ and $\tau_2(t)$ and finally calculating $\Gamma(t)$ from the last equation, analogous to the approach in Ref. \cite{Chen2012} which is based on Lewis-Riesenfeld invariants.
In Ref.  \cite{Chang} different equations for a  different model are considered. 

Basically, the PMP deals with finding an optimal control function $u^*(t)$ such that a given cost function $J$ of the form $J=\int_0^{t_1} L(u(t),...)dt$, where $L$ is a function of $u(t)$ and some state functions and their derivatives, is minimized for $u(t)=u^*(t)$.
Here, the time $T$ required for the protocol is to be minimized, $J=T$, and since one can write $T=\int_0^T1\,dt$ one has $L\equiv 1$. 

We first consider the case $\omega(t) \equiv \omega > 0$.
As the control function we choose $u(t)=\Gamma(t)$. The PMP then introduces the ``control Hamiltonian''
\beq \label{2.12}
H_c = -L + p_1 \dot\tau_1 + p_2 \dot\tau_2 + p_3 \dot\tau_3~,  ~~~~ L\equiv 1, 
\eeq
where $p_i=p_i(t)$ and where one inserts $\dot \tau_i$ from Eq. (\ref{2.9}), with $\Gamma$ replaced by $u$. Then $H_c$ assumes its maximum for $u=u^*$, the optimal control, and in addition one has $\dot p_i = -\partial H_c/\partial \tau_i$ when $u=u^*$.
Moreover, $H_c$ is constant along the optimal trajectory, and this constant is zero if the terminal time is free (i.e. not fixed), as in the present case.
In the following the asterisk on $u^*$ will be omitted. 

Since $u$ is unrestricted, the maximality of $H_c$ gives $\partial H_c / \partial u = 0$, and by Eq. (\ref{2.9}) this gives $\partial H_c / \partial u  =p_3=0$.
Then $\dot p_i = -\partial H_c/\partial \tau_i$ gives
\beq \label{dotp3}
\dot p_3= - \frac{\partial H_c}{\partial \tau_3} = 2\omega p_1 \sin \tau_3 + 2 \omega p_2\frac{\cos \tau_3}{\cos \tau_1} = 0
\eeq
and thus
\beqa
0 &=& 2\omega p_1 \sin \tau_3 + 2\omega p_2 \cos \tau_3/\cos\tau_1 \label{2.15}\\
\dot p_1 &=& -  \frac{\partial H_c}{\partial \tau_1} = 2\omega p_2 \sin\tau_3 \sin\tau_1/\cos ^2\tau_1\label{2.16}\\
\dot p_2 &=& - \frac{\partial H_c}{\partial \tau_2} =  0, ~~~~p_2= {\rm const}~\equiv~c_2.
\eeqa
Since $H_c \equiv 0$ for the optimal trajectory one obtains
\beq \label{2.18}
2\omega p_1 \cos\tau_3 - 2\omega c_2 \sin\tau_3/\cos\tau_1 = 1~.
\eeq
Multiplying Eq. (\ref{2.15}) by $\cos\tau_3$ and Eq. (\ref{2.18}) by $\sin\tau_3$ and then subtracting one obtains
\beq \label{2.19}
2\omega c_2/\cos\tau_1 = -\sin\tau_3~.
\eeq
Eqs. (\ref{2.15}) and (\ref{2.18}) then become 
$
\sin\tau_3\,(2\omega p_1 - \cos\tau_3) = 0$ and 
$\cos\tau_3\, (2\omega p_1 - \cos\tau_3)= 0$, 
and therefore
\beq
2\omega p_1= \cos\tau_3,~~~ \dot p_1 = -\dot\tau_3\sin\tau_3/2\omega~.
\eeq
With Eq. (\ref{2.19}) one obtains from Eq. (\ref{2.16}) for $\dot p_1$
\beq \label{dotp1}
\dot p_1 = - \sin\tau_3 \sin\tau_3 \sin\tau_1/\cos\tau_1.
\eeq
Inserting for $\dot\tau_3$ from Eq. (\ref{2.9}) one obtains $\Gamma \sin\tau_3=0$. Hence in any open interval in which $\Gamma \neq 0$ one has $\sin \tau_3 = 0$ and thus $\dot\tau_3 =0$, which then implies $\Gamma = 0$. 
Hence in the unconstrained case the optimal choice for $\Gamma$ is $\Gamma(t) \equiv 0$. Note that so far the initial and final state have not come into play.

If one allows an arbitrary $\omega(t)$, with $|\omega(t)| \leq \omega_{\rm max}$, one can introduce $\omega(t)$ as a second control function. An argument similar to the one above gives as optimal choice $ \Gamma(t) \equiv 0$ and $\omega(t) \equiv \pm \omega_{\rm max}$, where one can restrict oneself to the plus sign.

{\em Minimal time.} When $\Gamma(t)\equiv 0$ the time-development operator becomes $U_H(t,0)=\exp\{-i\omega\sigma_1t\}$ which in general does not satisfy Eq.~(\ref{2.4a}). Therefore one needs initial and final $\delta$ -like pulses of zero time duration (or, equivalently, initial and final conditions for $U_H$).
In the initial and final pulse, $\omega$ drops out when $|\Gamma| \to \infty$. The complete time-development operator for the optimal protocol from 0 to T is then
\beq \label{U_H}
U_H(T,0) = e^{-i\alpha_{\rm f}\sigma_3}\, e^{-i\omega \sigma_1T}\, e^{-i\alpha_{\rm in}\sigma_3}.
\eeq
For a given initial and final state one now has to determine all possible values of $\alpha_{\rm in,f}$ and $T$ such that Eq. (\ref{2.4a}) holds and then find the minimal $T$ among them \cite{referee}. 

To illustrate the procedure we consider as a specific example  the case where $|\psi_{\rm in}\ra$ is  the ground state of $H_{-\gamma}= -\gamma\sigma_3 + \omega \sigma_1$ and  $|\psi_f\ra$ the ground state of $H_\gamma= \gamma\sigma_3 + \omega \sigma_1$, with $\gamma >0$. The lowest eigenvalue of $H_{\pm\gamma}$ is given by 
$
\lambda_\gamma = - \sqrt{\gamma^2 + \omega^2}
$
and corresponding (non-normalized) eigenvectors $|\lambda_\gamma\ra_-$ and $|\lambda_\gamma\ra_+$  can be written as 
\beqa
|\lambda_\gamma\ra_- &=& \omega|0 \ra + ( \lambda_\gamma + \gamma)|1\ra\nonumber\\
|\lambda_\gamma\ra_+ &=& \omega|0 \ra +( \lambda_\gamma - \gamma)|1 \ra ~. \label{2.3}
\eeqa
With Eq. (\ref{2.3}) and 
$\exp(-i\omega\sigma_1 T) = \cos\omega T - i \sigma_1\sin\omega T
$
 one finds by means of a straightforward calculation
\begin{align}
&U_H(T,0)\,|\lambda_\gamma\ra _- \\
&= \{\omega \cos\omega T\,e^{-i(\alpha_{\rm in} + \alpha_{\rm f})}-i(\lambda_\gamma +\gamma)\sin\omega T\,e^{i(\alpha_{\rm in} - \alpha_{\rm f})}\}|0\ra\nonumber\\ &+ \{(\lambda_\gamma +\gamma)\cos\omega T\,e^{i(\alpha_{\rm in} + \alpha_{\rm f})}-i\omega\sin\omega T\,e^{-i(\alpha_{\rm in} - \alpha_{\rm f})} \}|1\ra.\nonumber
\end{align}
In order for this to equal a multiple of $|\lambda_\gamma\ra_+$ the ratios of the respective first and second component have to be equal, i.e.
\begin{align}
\frac{\omega \cos\omega T\,e^{-i(\alpha_{\rm in} + \alpha_{\rm f})}
-i(\lambda_\gamma +\gamma)\sin\omega\,T\,e^{i(\alpha_{\rm in} - \alpha_{\rm f})}}
{ (\lambda_\gamma +\gamma)\cos\omega T\,e^{i(\alpha_{\rm in} + \alpha_{\rm f})}
-i\omega\sin\omega\,T\,e^{-i(\alpha_{\rm in} - \alpha_{\rm f})} }\nonumber\\
=\frac{\omega }{\lambda_\gamma -\gamma}
 \label{2.26a}
\end{align}
which by a simple calculation gives
\begin{align}
0 =&~ \omega\gamma \cos\omega T(e^{i(\alpha_{\rm in} + \alpha_{\rm f})} + e^{-i(\alpha_{\rm in} + \alpha_{\rm f})} )\nonumber\\
&+\omega^2\sin\omega T\,i(e^{i(\alpha_{\rm in} - \alpha_{\rm f})}-e^{-i(\alpha_{\rm in} - \alpha_{\rm f})} \}\nonumber\\
&+\omega\lambda_\gamma \cos\omega T\,(e^{i(\alpha_{\rm in} - \alpha_{\rm f})} - e^{-i(\alpha_{\rm in} - \alpha_{\rm f})} )
\end{align}
The imaginary part (in the last line) gives $\sin(\alpha_{\rm in} + \alpha_{\rm f})=0$, i.e. $\alpha_{\rm f} = - \alpha_{\rm in}+ n\pi$, and with this the real part gives 
\beq
2\gamma \cos\omega T e^{in\pi} = 2\omega \sin\omega T e^{in\pi}\sin2\alpha_{\rm in}
\eeq
Hence $\tan\omega T = \gamma/\omega\sin2\alpha_{\rm in}$ and 
the time duration $T$ becomes minimal for $\alpha_{\rm in,f}=\pm\pi/4$. Thus one obtains
\beq \label{2.26b}
\tan \omega T_{\rm min} = \gamma/\omega~.
\eeq
For $\gamma/\omega =2$ this agrees with the experimental result of Ref. \cite{Mo} and also with the simulations.

 For general initial and final (normalized) states
$|\psi_{\rm in}\ra = i_0|0\ra + i_1|1\ra $ and
$|\psi_{\rm f}\ra = f_0|0\ra + f_1|1\ra~$, where
 $\sigma_3|0,1\ra =\pm |0,1\ra$,  one obtains by the same procedure the amazingly simple result
\beq \label{cos}
\cos \omega T_{\rm min} = |f_0i_0| + |f_1 i_1|~.
\eeq
If also  $\omega(t)$ is time-dependent, with $|\omega(t)| \leq \omega_{\rm max}$, the minimal time is obtained by replacing $\omega$ by $\omega_{\rm max}$ in Eq.~(\ref{cos}). For optical driving of a two-level system, with fixed detuning $\Delta$ and the Rabi frequency $\Omega(t)$ as an unconstrained control, $\omega$ in Eq. (\ref{cos}) is replaced by $\Delta$ and the components $i_k, f_k$ by the components with respect to the eigenvectors of $\sigma_1$.
Note that $T_{\rm min}$ is symmetric under the interchange of initial and final state. Also, $T_{\rm min}=0$ if and only if $|f_1||i_0|=|f_0||i_1|$, i.e. $|f_j|=\mu |i_j|$ for some  $\mu >0$.

The initial and final $\delta$ pulse depend in general on the relative phases of $i_0$ and $i_1$ and of $f_0$ and $f_1$ and on the ratios of the components.
 In particular, if $|\psi_{\rm in}\ra$ and $|\psi_{\rm f}\ra$ are the  groundstate of  $H_{\gamma_{\rm in}}$ and $H_{\gamma_{\rm f}}$, respectively, then one can choose $\alpha_{\rm f}=-\alpha_{\rm in}$ and $\alpha_{\rm in} = \pi/4$ for $\gamma_{\rm in}< \gamma_{\rm f}$ and $\alpha_{\rm in} = -\pi/4$ for  $\gamma_{\rm in}> \gamma_{\rm f}$.

{\em Constrained driving.}  Let $|\Gamma(t)|\equiv u(t) \leq c$. 
From Eqs. (\ref{2.9}) and (\ref{2.12}), the only term in $H_c$  which contains $u$ is of the form $2 p_3 u$. If a maxixum of $H_c$ is reached for $u$ lying in the interior of the interval [-c,c] then $\partial H_c /\partial u = 0$, and then $\Gamma(t) \equiv 0$, as before. If it lies on the boundary then $u=c$ or $u=-c$. Therefore, since  one expects that the initial and final $\delta$-pulses are replaced by a time development of finite duration,  we make an ansatz with an initial period of length $T_c\geq 0$ with $\Gamma(t)= c$, then a period of length $T_{\rm off}\geq  0$ with $\Gamma(t)=0$, and a final period of length $T_{-c}\geq 0$ with $\Gamma(t)=-c$, where some of the three times may be zero. The total time $T = T_c + T_{\rm off} + T_{-c}$ should be minimal, with all three terms non-negative. 

The Hamiltonians in the three respective time periods are $\pm c\sigma_3+\omega \sigma_1$ and $\omega\sigma_1$ which gives
\beq 
U_H(T,0)=e^{-i(- c\sigma_3+\omega \sigma_1) T_{-c}} e^{-i\omega\sigma_1 T_{\rm off}} e^{-i ( c\sigma_3+\omega \sigma_1) T_c}  \nonumber
\eeq
Here we consider only initial and final states $|\lambda_\gamma\ra_\pm$. The same procedure as above then shows
 that $T_{-c}=T_c$ and that $T_{\rm off}$ can be expressed as a function of $T_c$ so that the total time $T$ becomes a function of $T_c$. This latter function has to be minimized. 
For $c \geq \omega^2/\gamma $ one obtains $T_{\rm min}= 2T_c+T_{\rm off}$ with
\begin{align}
T_c& = \frac{1}{\sqrt{c^2+\omega^2}}\arcsin\sqrt{\frac{c^2+\omega^2}{2c (c+\gamma)}}\label{3.5c}\\
T_{\rm off}&= \frac{1}{\omega}\arctan\frac{c\gamma-\omega^2}{\omega\sqrt{c^2+2c\gamma-\omega^2}}\label{3.5d}
\end{align}
For $c \to \infty$ Eq.~(\ref{3.5d}) becomes Eq.~(\ref{2.26b}) of the unconstrained case. Furthermore,  as  $c \to \infty$, $T_c \to 0$ and $(c^2+\omega^2)^{1/2} T_{\rm c} \to \pi/4 $ so that the initial and final periods approach a $\delta$ pulse in $ \sigma_3$ of strength $\pm\pi/4$, as in the unconstrained case.  For $ c \leq \omega^2/\gamma $ one has $T_{\rm off}= 0$ and
\begin{align}
T_{\rm min}(c) &=2T_c= \frac{2}{\sqrt{c^2+\omega^2}}\arcsin\sqrt{\frac{\gamma(c^2+\omega^2)}{2\omega^2(c+\gamma)}}~.
\end{align}
Thus for $c > \omega^2/\gamma$ and for the states considered here, the optimal protocol is of bang-off-bang type, while for $c \leq \omega^2/\gamma $ it is bang-bang.

If there is a switching time $\epsilon>0$ for $\Gamma$ from $c$ to $0$ and  0 to $-c$, with $\omega\epsilon,\,c\epsilon \ll 1$, and if one retains $T_c$ and $T_{\rm off}$ from above, then for the fidelity ${\cal F}$ one has ${\cal F} > 1-2(\omega \epsilon + c\epsilon)$, instead of 1. This bound is 
 {\em independent} of the shape of the switching function. Finite coherence times much longer than $T_{\rm min}$ also have only a small effect on $\cal F$. Instead of keeping $T_c$ and $T_{\rm off}$ from Eqs. (\ref{3.5c},\ref{3.5d}) one can change them
to increase the fidelity to 1, up to terms of second order in $\omega
\epsilon$ and $c \epsilon$. E.g., for a linear switch pulse one just
has to use $T_c - \epsilon/2$  and $T_{\rm off} - \epsilon$.

\begin{figure}[tb]
\begin{center}
\includegraphics[width=.5\textwidth]{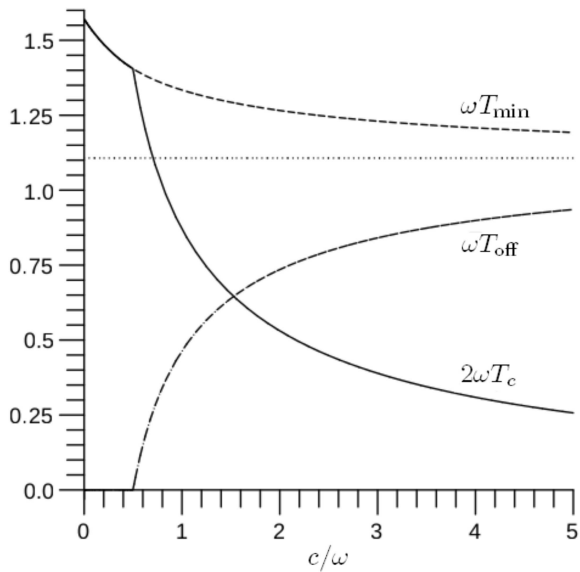}
\caption{$\omega T_{\rm min}$,  $\omega T_{\rm off}$ and $2\omega T_c$, the double of the corresponding bang duration $\omega T_c$, as a function of $c/\omega$
 for  $\gamma/\omega = 2$. For $c\to \infty$ one sees that $\omega T_{\rm min}$ approaches the unconstrained value $ \arctan \gamma/\omega$.
 For $c/\omega \leq \omega/\gamma = 0.5$ there is no period with $\Gamma(t)\equiv 0$, i.e $T_{\rm off} = 0$, so that for these values of $c$ the protocol is of bang-bang type. }
\label{Tmin}
\end{center}
\end{figure}

In Fig. \ref{Tmin},  $\omega T_{\rm min}$ is plotted as a function of $c/\omega$ for $\gamma/\omega = 2$,  the off duration $T_{\rm off}$, the asymptote $\arctan \gamma/\omega$ for the unconstrained case and $2 T_c$, the double of the corresponding individual bang duration.

{\em Quantum speed limit time.} For a time-independent Hamiltonian the state overlap $|\la\psi_t|\psi_0\ra|$ is bounded by the Fleming-Bhattacharyya bound \cite{FleBhat}, $t \geq \arccos|\la\psi_0|\psi_t\ra|/(\Delta E/\hbar)$,  with $\Delta E$ the energy variance  of the state. For the time-dependent LZ Hamiltonian in Eq. (\ref{2.1})  and initial and final state momentary eigenstates of the LZ Hamiltonian, Ref. \cite{Mo} used a quantum speed limit time $T_{\rm qsl}$,  with $\cos \omega T_{\rm qsl} =|\la \psi_{\rm f}|\psi_{\rm in}\ra|$,
while Ref. \cite{Caneva} suggested an expression with $\omega$ replaced by
$\Delta E_0/\hbar $, the energy variance of the initial state with respect to $H_0=\omega \sigma_1$.  Since $\Delta E_0/\hbar$ is strictly smaller than $\omega$, the quantum speed limit time used in Ref. \cite{Caneva} is larger than $T_{\rm qsl}$ of Ref. \cite{Mo}. 
Using Eq. (\ref{cos}) one finds
\begin{align}
\cos\omega T_{\rm qsl} =  |\bar{f}_0i_0+\bar{f}_1i_1|
                     \leq |f_0i_0|+|f_1i_1|
                    =\cos\omega T_{\rm min}\label{estim}
\end{align}
and since the cosine is strictly decreasing in $[0,\pi/2]$ this shows that one always has $T_{\rm min}\leq T_{\rm qsl}$.

Moreover, one has $T_{\rm min} = T_{\rm qsl}$ if and only if the relative phase of $f_0$ and $f_1$ and that of $i_0$ and $i_1$ are equal since this is the condition for the equality sign to hold  in the estimate in Eq. (\ref{estim}).
This is true in particular for groundstates of $H_{\gamma_{\rm in}}$  and
 $H_{\gamma_{\rm f}}$, but not if one is a groundstate and the other the excited state. For example, if $|\psi_{\rm in}\ra$ and $|\psi_{\rm f}\ra$ are the groundstate and excited state of $H_\gamma$, for a fixed $\gamma$, then $T_{\rm qsl}=\pi/2\omega$, while $T_{\rm min}=1/\sqrt{\gamma^2+\omega^2}$, which is less than $T_{\rm qsl}$.

{\em Discussion.}
We have shown above that the speed limit time $T_{\rm qsl}$ used in Ref.\cite{Mo} coincides with $T_{\rm min}$ provided one considers only momentary eigenstates of the LZ Hamiltonian. However,  for more general initial and final states the time $T_{\rm min}$ from Eq. (\ref{cos}) for the optimal protocol may actually be {\em smaller} than $T_{\rm qsl}$. Therefore, for more general initial and final states the particular expression for $T_{\rm qsl}$ in Ref. \cite{Mo} does not seem to be appropriate, and neither does that of  Ref. \cite{Caneva} since it is always larger than the former. A more natural  choice as a quantum speed limit time seems to be  $T_{\rm min}$. 

It  should be noted that $T_{\rm min}$ goes to zero for increasing $\omega$ and 
therefore  a universal quantum speed limit time would be zero. Hence a meaningful quantum speed limit time should only be defined with respect to a given class of available Hamiltonians, such as above for the LZ type Hamiltonians. This is exemplified in Fig. \ref{Tmin}, where the minimal time depends on the strength of the available driving.
It would be interesting to find an expression that applies to a  Hamiltonian class  as large as possible.

 I would like to  thank O. Morsch and J.G. Muga  for stimulating discussions.


\begin{thebibliography}{99}
\bibitem{referee} Alternatively, this can also be done on the Bloch sphere by writing Eq. (\ref{U_H}) in terms of rotations, as pointed out by a referee. The computational efforts are comparable.
\end{thebibliography}
\end{document}